\documentclass[12pt]{article}

\usepackage{cite}
\usepackage{epsfig}
\usepackage{amsmath}
\usepackage{pstricks}
\usepackage{bbold}

\def\mathswitch#1{\relax\ifmmode#1\else$#1$\fi}
\def\mathswitchr#1{\relax\ifmmode{\mathrm{#1}}\else$\mathrm{#1}$\fi}

\begin{document}
\thispagestyle{empty}

\def\thefootnote{\fnsymbol{footnote}}

\begin{flushright}
CP3--10--35
\end{flushright}

\vspace{1cm}

\begin{center}

{\Large\sc {\bf The Matrix Element Method and QCD Radiation}}
\\[3.5em]
{\large\sc
J.~Alwall$^1$, A.~Freitas$^2$, O.~Mattelaer$^3$
}

\vspace*{1cm}

{\sl $^1$
Department of Physics and National Center for Theoretical Sciences, \\
National Taiwan University, Taipei, 10617, Taiwan 
}
\\[1em]
{\sl $^2$
Department of Physics \& Astronomy, University of Pittsburgh,\\
3941 O'Hara St, Pittsburgh, PA 15260, USA
}
\\[1em]
{\sl $^3$
Center for Cosmology, Particle Physics and Phenomenology (CP3), Universit\'e Catholique
de Louvain, Chemin du Cyclotron 2, 1348 Louvain-la-Neuve, Belgium
}

\end{center}

\vspace*{2.5cm}

\begin{abstract}
The matrix element method (MEM) has been extensively used for the
analysis of top quark and W-boson physics at the Tevatron, but in general
without dedicated treatment of initial-state QCD radiation. At the LHC, the
increased center-of-mass energy
leads to a significant increase in the amount of QCD radiation, which
makes it mandatory to carefully account for its effects. We here
present several methods for inclusion of QCD radiation effects in the
MEM, and apply them to mass determination in the presence of
multiple invisible particles in the final state. We demonstrate
significantly improved results compared to the standard
treatment.
\end{abstract}

\setcounter{page}{0}
\setcounter{footnote}{0}

\newpage


\section{Introduction}

The discovery and analysis of new physics particles, beyond the
Standard Model, is one of the main goals for building the Large
Hadron Collider (LHC). The presence of such new physics at an energy
scale accessible by the LHC is indicated by shortcomings of the
Standard Model, in particular its inability to explain the stability
of the Higgs boson mass and the nature of the cosmological dark
matter. New physics models which simultaneously explain these two
riddles have certain common features: strongly produced particles
charged under QCD, which are perhaps exclusively pair-produced and
decay (mainly or exclusively) to stable invisible particles ($i.\,e.$,
particles with negligible interactions with the detector
material). Even if the new physics includes singly produced
resonances, these might very well decay mainly to final states
including invisible particles.

The analysis of new physics processes with invisible particles in the
final state is particularly challenging, due to the absence of
invariant mass peaks. Even the determination of masses of the new
particles typically requires large statistics
\cite{mass_determination_with_missing_energy}. This difficulty is
further enhanced by the presence of QCD radiation (in particular
initial-state radiation, $i.\,e.$ QCD radiation that is not collinear with
any of the decay products), which complicates the final states in
several ways: by adding additional jets (that do not correspond to
any decay products of the produced new physics particles), and by
modifying the kinematics of the decay products through emission of
particles with sizeable transverse momenta, which result in
transverse boosts of the decay products.

Among the methods for extraction of model parameters, the one which
takes into account the most information from the experimental events
is the Matrix Element Method (MEM) \cite{matrix}. The MEM has been
used extensively for top quark physics by the experimental
collaborations at the Tevatron \cite{toptev}, and has resulted in
the most precise measurement of the top quark mass to date. The
method is described in detail in section~\ref{sc:method}, and it here
suffices to say that it uses the theoretical squared matrix
element for the process to give a weight for each experimental event,
for a given point in the parameter space of the model. One drawback
of the method in its basic form
is that only events with final states that perfectly
match those in the considered matrix element can be used. While this
is not a big problem at the Tevatron, where the phase space for
additional QCD radiation is limited by the relatively low
center-of-mass energy, events at the LHC have a large phase space for
QCD radiation. It is therefore of crucial importance to develop the
MEM to account also for such additional radiation.

In this paper, we suggest several methods to account for additional
QCD radiation in processes with invisible particles in the final
state, either by explicitly correcting for the QCD emission
or by parametrizing its effect through an appropriate function.
Partial results have already been reported in Ref.~\cite{susy}, but 
this paper provides a more extensive and detailed analysis.
It is organized as follows: In section~\ref{sc:method},
we describe the details of the MEM. Section~\ref{sc:parton} is
dedicated to the study of the MEM with initial-state QCD radiation at
the parton level, using the parton shower Monte Carlo generator
\textsc{Pythia} \cite{pythia}. We study two example processes, top quark pair
production with di-leptonic decay giving neutrinos as decay products
for both top quarks, and Standard Model Higgs boson production with
decay to $W^{+}W^{-} \to l^+l^-\nu\bar\nu$. Although these are
both Standard Model processes, they are representative for the type of
new physics we are targeting---heavy particles with invisible
particles in their decays, which makes it impossible to reconstruct
the complete final state of any of the produced particles.  In
section~\ref{sc:hadron}, we turn to the more complicated case where
both QCD radiation and hadronization as well as detector effects are
taken into account. We give our conclusions in section~\ref{sc:concl}.

Note that similar approaches to include QCD radiation have been explored 
by the D0 and CDF collaborations in Refs.~\cite{tevisr}.


\section{The Matrix Element Likelihood Method}
\label{sc:method}

The Matrix Element Method (MEM) \cite{matrix} provides a recipe for
computing a likelihood that an experimental event agrees with a
theoretical model. The information about the theoretical model is
supplied in the form of the squared matrix element for a concrete
process. One or several parameters of the model can be determined from
the data by finding the maximum of the likelihood for a sample of
events as a function of the parameters.

The likelihood measure for a single event, with measured momenta
$\textbf{p}_i^{\rm vis}$, to agree with the model, given a set of parameters
$\alpha$, is defined as
\begin{align} 
{\cal P}(\textbf{p}^{\rm vis}_i|\alpha) &= \frac{1}{\sigma_\alpha} \int dx_1 dx_2 \, 
 \frac{f_1(x_1)f_2(x_2)}{2sx_1x_2}
\left[ \prod_{i \in \text{final}} \int 
 \frac{d^3 p_i}{(2\pi)^3 2E_i} \right] |M_\alpha(p_i)|^2
 \prod_{i \in \text{vis}} \delta(\textbf{p}_i - \textbf{p}^{\rm vis}_i).
 \label{mem1}
\end{align}
Here $f_1$ and $f_2$ are the parton distribution functions,
$M_\alpha$ is the theoretical matrix element, 
and $\sigma_\alpha$ is the total cross section, computed with the same matrix
element. The normalization ensures that 
\begin{align}
\left[ \prod_{i \in \text{vis}}\int d^3p_i^{\rm vis} \right]
 {\cal P}(\textbf{p}^{\rm vis}_{i}|\alpha)=1.
\end{align}
The three-momenta $\textbf{p}_i^{\rm vis}$ of the visible measured objects 
are matched with the corresponding momenta $\textbf{p}_i$ of the final-state particles in the
matrix elements, while the momenta of invisible particles (neutrinos or weakly
interacting new physics particles) are integrated over.

Quarks and gluons are matched with jets as visible objects, whose energy is
typically not measured very precisely. Therefore one has to include a transfer
function $W$ for jets, which parametrizes the combined effects of parton
showering, hadronization and detector response:
\begin{align} 
{\cal P}(\textbf{p}^{\rm vis}_i|\alpha) &= \frac{1}{\sigma_\alpha} \int dx_1 dx_2 \, 
 \frac{f_1(x_1)f_2(x_2)}{2sx_1x_2}
\left[ \prod_{i \in \text{final}} \int 
 \frac{d^3 p_i}{(2\pi)^3 2E_i} \right] |M_\alpha(p_i)|^2
 \prod_{i \in \text{vis}} W_i(\textbf{p}_i,\textbf{p}^{\rm vis}_i).
 \label{mem2}
\end{align}
For leptons and photons it is mostly sufficient to approximate the transfer
functions by a delta function, as above.

For a sample of $N$ events, the combined likelihood is usually stated in terms
of its logarithm:
\begin{align}
-\ln({\cal L}) &= -\sum_{n=1}^N \ln {\cal P}(\textbf{p}^{\rm vis}_{n,i}|\alpha)
 + N \left[ \prod_{i \in \text{vis}} \int d^3p_i^{\rm vis} \right]
 {\rm Acc}(\textbf{p}^{\rm vis}_{i}) \, {\cal P}(\textbf{p}^{\rm vis}_{i}|\alpha),
\label{logl}
\end{align}
where $\textbf{p}^{\rm vis}_{n,i}$ are the measured momenta of the $n$th
event.\footnote{In the likelihood, we have neglected the term  $\sum_{n=1}^{N}
\ln({\rm Acc}(\textbf{p}^{\rm vis}_{n,i}))$ since it is independent of any theoretical
parameters $\alpha$.} The acceptance function ${\rm Acc}(\textbf{p}^{\rm
vis}_{i})$ corrects for the bias introduced by detector acceptance and event
selection\footnote{When using selection cuts pertaining to individual jets, the
likelihood can be shifted by a small bias even when the acceptance
function is properly incorporated in the calculation. This is related to the fact that
the transfer functions $W_i$ only imperfectly model the effect of extra jets
from final-state radiation. For the content of this paper this bias is
irrelevant, since we mainly focus on the role of initial-state radiation, but
let us remark that in principle one could construct an un-biased estimator with
a more sophisticated parametrization of the transfer
functions.\label{fsrtrans}}.\label{fsrtrans2}

\vspace{\medskipamount}
The MEM has been used extensively for top quark physics by the
experimental collaborations at the Tevatron \cite{toptev}. At this
time, the most precise determination of the top quark mass has been
achieved with this technique, which can be attributed to the fact that
in principle it takes into account all relevant experimental information. In
most Tevatron analyses\footnote{See Ref.~\cite{tevisr} for approaches 
including extra jets.}, only events for which the number of jets
exactly matches the number of colored partons in the hard matrix
element have been included. For instance, only two-jet events have
been considered for the analysis of the di-leptonic top quark pairs.
At the Tevatron this approach works, since the top quarks are
relatively heavy compared to the beam energy and thus the phase space
for extra radiation is highly suppressed.

However, at the LHC radiation of hard jets is expected to be abundant, not only
for top pair production but also for new physics processes involving colored
particles with masses of a few 100 GeV \cite{sjet}.  With a center-of-mass
energy of 14~TeV, the number of top pair events is reduced by more than 40\%  if
a cut on extra jets with $p_{\rm T} > 40$~GeV is imposed\footnote{About 27\% of
events have one additional jet with $p_{\rm T} > 40$~GeV, while about 15\% have
two or more extra jets.}. This estimate is based on events generated with
\textsc{Pythia 6.4} \cite{pythia} and passed through the fast detector
simulation \textsc{Pgs 4} \cite{pgs}. As will be shown later, even the presence
of additional jets with $p_{\rm T} < 40$~GeV can lead to problems with fitting
the signal events, so that a tighter cut will be necessary to
sufficiently reduce the
influence of jet radiation, hence leading to a large loss of signal statistics.

Alternatively, one could try to take into account events with extra jets by
including matrix elements with more partons in the final state. Referring again
to the example of di-leptonic top quark pairs, this would amount to matrix elements
corresponding to the processes $pp \to b\bar{b} l^+l^-\nu\bar{\nu} + n_qq +
n_{\bar{q}}\bar{q} + n_gg$, which have $n_q$ quarks, $n_{\bar{q}}$ antiquarks
and $n_g$ gluons in the final state besides the usual top decay products.
While this approach should allow to correctly include all events, it
substantially increases the computation time, due to the complexity of the
multi-particle matrix elements, the more complicated structure of the phase space,
and the combinatorics related to summing over 
quark flavors and gluons in the extra jets. Even if one restricts oneself to
considering only up to one extra parton in the final state, the computing
intensity of
the likelihood fit is increased by more than one order of magnitude.

The large majority of extra
jets originates from initial-state radiation (ISR). In the following two
sections, several methods will be
described which account for the main effect of ISR by performing
kinematical corrections event by event, using matrix elements for the hard
process only, without additional partons in the final state.

\vspace{\medskipamount}
For concreteness, the numerical analyses in the following sections has been
carried out for two representative processes with invisible particles in the
decay, so that the final state is not fully reconstructable. 
Top quark pair production with di-leptonic decay,
\begin{align}
pp \to t\bar{t} \to b\bar{b} l^+l^{\prime-}\nu_l\bar{\nu}_{l'}, 
\label{proc}
\end{align} 
is a typical case of pair production of heavy particles with
relatively long and not fully reconstructable decay chains. As a second example
we will consider Higgs production via gluon fusion,
\begin{align}
pp \to gg \to h \to W^+W^- \to l^+l^{\prime-}\nu_l\bar{\nu}_{l'}, 
\label{proc2}
\end{align} 
with the characteristic features of a $s$-channel resonance.

Numerical results shown in the following sections correspond to a center-of-mass
energy of $\sqrt{s}=14$~TeV, but the essential aspects do not change
for lower values of $\sqrt{s}$.
Two independent implementations of the MEM are employed: the first is a
specialized private
code written by us using matrix elements generated by \textsc{CompHEP 4.4}
\cite{comphep}, the second is the flexible automated public tool
\textsc{MadWeight} \cite{madweight}, which is based on the
\textsc{MadGraph/MadEvent} framework \cite{ME}. 

The core task of both programs is the evaluation of the integration in
eqs.~\eqref{mem1}, \eqref{mem2}. For the process \eqref{proc} this involves eight integration
variables (six momentum components for the neutrino and the
anti-neutrino, plus $x_1$ and $x_2$), but energy-momentum conservation reduces it
to a four-dimensional integration.
Instead of the final-state (anti-)neutrino momenta, we use the invariant masses of the top quarks and W bosons as
integration variables, since the integrand has sharp resonance peaks as function of 
these variables. The transformation between the neutrino momenta and the
intermediate invariant masses involves the solution of a coupled system of two
quadratic equations, which can be performed analytically. The
final expressions are rather long and will therefore not be shown here 
(see Ref.~\cite{madweight} for examples).
The Breit-Wigner resonances of the form
\begin{align}
\frac{1}{(q_k^2-m_k^2)^2+m_k^2\Gamma_k^2}, && (k=t,\bar{t},W^+,W^-),
\end{align}
in the integrand can be mapped out by the additional
variable transformation
\begin{align}
q_k^2 &= m_k^2 + m_k\Gamma_k \tan t_k, &
\frac{d(q_k^2)}{dt_k} &= \frac{(q_k^2-m_k^2)^2+m_k^2\Gamma_k^2}{\pi
 m_k\Gamma_k},
\end{align}
so that as a function of the $t_k$ 
the integrand becomes almost flat across the entire region $-\pi/2 \leq
t_k \leq \pi/2$ and the integration can be performed numerically without
difficulty. The outputs of the two programs
agree very well for all results shown in the following sections.

\begin{figure}
\centering
\epsfig{figure=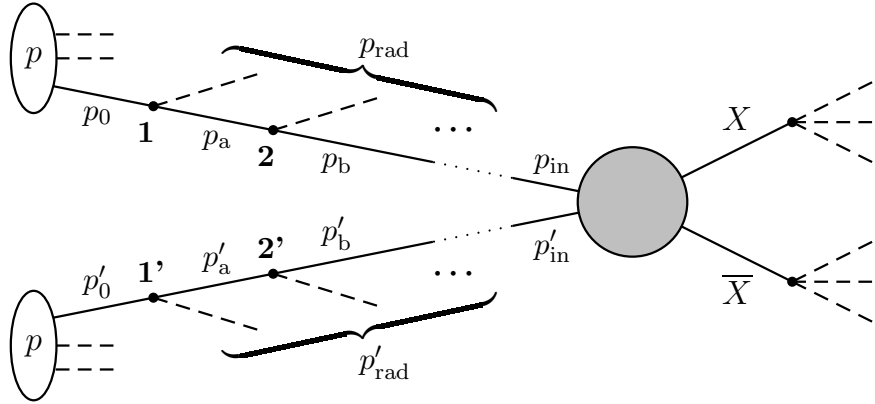, width=11.5cm}
\caption{Schematic depiction of the event topology for pair production of heavy
particles $X$, together with initial-state radiation.}
\label{diag}
\end{figure}


\section{Initial-State Radiation at Parton Level}
\label{sc:parton}
 
In this section we restrict ourselves to an analysis at the parton level.
Simulated ``data'' events have been generated with \textsc{Pythia 6.4}
\cite{pythia} using CTEQ6L1 parton distribution functions \cite{cteq}.
The momenta of the final-state particles in \eqref{proc} or \eqref{proc2}, as
well as those of the initial-state radiation, have
been extracted from the \textsc{Pythia} event record for each event.
No cuts on the parton momenta have been implemented and therefore the acceptance
term in \eqref{logl} is simply 1.
This allows us to single out the effects of ISR without
complications from final-state radiation, hadronization, underlying
event and detector efficiencies.
For simplicity and clarity of the discussion, we do not include backgrounds in
the analysis.

The first technique that we propose here is based on the
observation that the most significant effect of ISR is on the
kinematics of the events; without proper inclusion of ISR the momentum
balance of the decay products is violated. The proper kinematics of
the hard scattering matrix element can be restored by simply boosting
the hard event by the momenta of the ISR. Since the longitudinal
incoming momenta are integrated over in the computation of the
likelihood, see eqs.~\eqref{mem1} and \eqref{mem2}, it is sufficient
to perform the boost for the transverse coordinates only.  In
practice, instead of boosting the measured final-state momenta, we
perform the boost on the incoming partons of the matrix element, which
is equivalent since the squared matrix element is a Lorentz scalar.
With this technique we are only performing a kinematical boost, which
allows us to sum up the ISR momenta for each incoming leg---the
sequence of individual branchings does not play any r\^ole.

This boost correction is the simplest possible treatment of ISR, which only
maintains the proper momentum balance, while the effects of the
particular QCD vertices and internal propagators (labeled by numbers and
$p_{a,b,...}$
in Fig.~\ref{diag}, respectively) are not taken into account.
It has the advantage of not increasing the computing time of the MEM likelihood
fit compared to the situation without ISR.

However, one can try to do better by including Sudakov reweighting for the ISR.
The Sudakov factor corresponds to the probability for no branching to occur between two
scales $p^2_{{\rm T},E1} < p^2_{{\rm T},E0}$. For ISR it is appropriate to
formulate the Sudakov factor in terms of
backwards evolution from the hard process to the incident proton.
In this case it is given by
\begin{align}
&\begin{aligned}
\Delta_{\rm ISR}&(p^2_{{\rm T},E0}, p^2_{{\rm T},E1}) \\
&\!\!=\exp \biggl ( -\int_{p^2_{{\rm T},E1}}^{p^2_{{\rm T},E0}} 
 \frac{d(p^2_{{\rm T},E})}{p^2_{{\rm T},E}}\,
 \frac{\alpha_{\rm s}(p^2_{{\rm T},E})}{2\pi} 
 \sum_{j\in \{ j\to i+X \}} 
 \int_{z_{\rm min}(p^2_{{\rm T},E})}^{z_{\rm max}(p^2_{{\rm T},E})} 
 dz \frac{P_{j\to i}(z)}{z}\,
 \frac{f_j(x_i/z,p^2_{{\rm T},E})}{f_i(x_i,p^2_{{\rm T},E})} 
\biggr )
\end{aligned}
\end{align}
where the sum runs over all possible assignments of partons $i,j$ (quarks or
gluon) in the branching $j\to i+X$. Here $P_{j\to i}$ are the splitting
functions, which for massless quarks read 
\begin{align}
P_{qq}(z) &= P_{qg} = \frac{4(1+z^2)}{3(1-z)}, &
P_{gq}(z) &= \frac{1}{2}\left[ z^2+(1-z)^2\right], &
P_{gg}(z) &= 6\frac{[1-z(1-z)]^2}{z(1-z)}.
\end{align}
Furthermore, $z$ is the ratio between the
pre-branching invariant mass squared of the parton-parton interaction and the
post-branching invariant mass squared.

To account for the proper weight of the ISR,
one needs
the probability of having a splitting $j\to i+X$ at some kinematic configuration
$(p^2_{{\rm T},E},z)$, which is
given by taking the derivative of the Sudakov factor:
\begin{align} 
{\cal P}_j(p^2_{{\rm T},E},z) 
&= -\frac{d^2}{d(p^2_{{\rm T},E})dz} 
 \Delta_{\rm ISR}(p^2_{{\rm T},E0}, p^2_{{\rm T},E}) \\
&= \frac{\alpha_{\rm s}(p^2_{{\rm T},E})}{2\pi p^2_{{\rm T},E}} 
 \,\frac{P_{j\to i}(z)}{z}\, \frac{f_j(x_i/z,p^2_{{\rm T},E})}{f_i(x_i,p^2_{{\rm T},E})}
 \Delta_{\rm ISR}(p^2_{{\rm T},E0}, p^2_{{\rm T},E})
\label{sud}
\end{align}
The branching probability for any kind of parton
is then given by $\sum_j {\cal P}_j(p^2_{{\rm T},E},z)$.

Evidently it is not possible to reconstruct the entire sequence of ISR
branchings in the correct order from the event data. For most events, however,
one microscopic branching process carries most $p_{\rm T}$ of all ISR from one
leg, so that a reasonable approximation can be obtained by adding up all ISR
momenta stemming from one leg and calculating the Sudakov factor for one single
branching with the summed momentum $p^2_{{\rm T},E} = p^2_{{\rm T,ISR}}$.

Since the ISR tends to be emitted at low angles, we approximate the ratio $z$ by
the longitudinal momentum components,
\begin{align}
z \approx \frac{p_{\rm in, z}}{p_{\rm in,z}+p_{\rm rad,z}},
\end{align}
where $p_{\rm in}$ is the momentum of the incoming parton of the hard collision
process and $p_{\rm rad}$ is the momentum of the ISR associated with this leg,
see Fig.~\ref{diag}.

\begin{figure}
\centering
\epsfig{figure=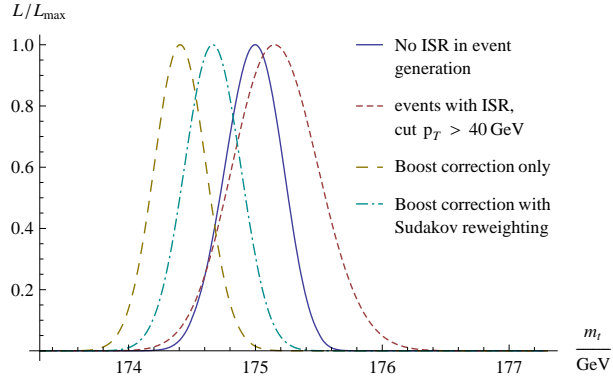, width=8cm}
\caption{Reconstruction of the top quark mass from a matrix
  element likelihood fit to 1000 parton-level di-lepton $t\bar{t}$ events at the LHC with
  $\sqrt{s} = 14$~TeV. A top mass of $m_t = 175$~GeV has been used for the event
  generation. Shown in the plot are the idealized situation without ISR in the
  event generation (solid curve), the influence of ISR if no correction
  method is used (short dashed), the result for application of the kinematic boost
  correction only (long dashed), and the boost correction with Sudakov reweighting
  (dash-dotted).
  The likelihood reflects statistical errors only.}
\label{parton}
\end{figure}
\vspace{\medskipamount}
Fig.~\ref{parton} shows numerical results for the MEM likelihood fit for the
example of top quark pair production with di-leptonic decays, eq.~\eqref{proc}.
For reference, the solid curve shows the idealized situation without ISR in
the event generation, so that the events are directly produced by the same
matrix element that is used in the matrix element analysis, and
contain exactly two $b$-quarks and two leptons.
The short-dashed curve corresponds to
event generation with ISR, but ISR is not accounted for in the
likelihood fit; instead events with ISR with combined $p_{\rm T} > 40$~GeV have been
vetoed. In this (parton-level) case, the fit 
yields a central value for the top quark mass close to the true input value,
but the statistical uncertainty is increased by a factor of about 1.5
(as can be seen from the larger width of the curve).

The other curves in Fig.~\ref{parton} demonstrate the effect of the boost
correction. As expected, the statistical error is not increased by this
method. Applying only the kinematical boost correction, without the Sudakov
reweighting, leads to a central value for the fitted top mass that is shifted
downwards by about 0.5~GeV. While this is still marginally consistent within
errors, it is indicative of a slight bias.  If in addition the Sudakov factor
\eqref{sud} is included, the central value of the reconstructed top mass is much
closer to  the true input value and fully consistent within errors.

\begin{figure}
\centering
\epsfig{figure=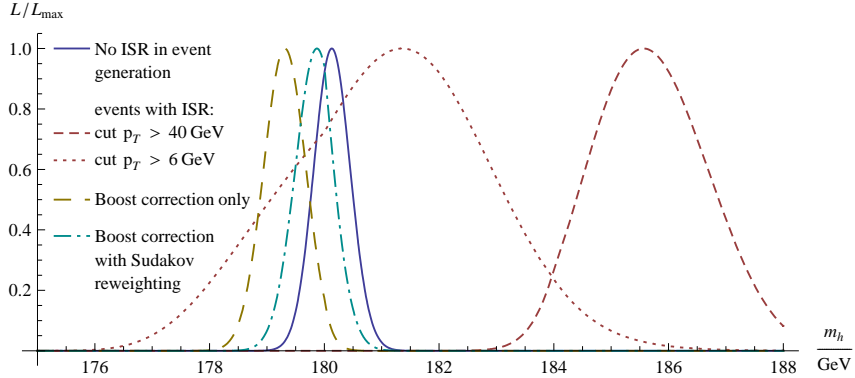, width=11.25cm}
\caption{Reconstruction of the Higgs boson mass from a matrix
  element likelihood fit to 1000 parton-level di-lepton events at the LHC with
  $\sqrt{s} = 14$~TeV. A Higgs mass of $m_h = 180$~GeV has been used for the event
  generation, so that the $W^+W^-$ channel is open and dominant.
  The plot compares the idealized situation without ISR in
  the event generation (solid curve) with simulation results including ISR, 
  and either no correction (short
  dashed and dotted), purely kinematical boost correction (long dashed),
  or boost correction with Sudakov reweighting (dash-dotted).
  The likelihood reflects statistical errors only.}
\label{parton2}
\end{figure}
\vspace{\medskipamount}
Fig.~\ref{parton2} shows
results for Higgs production with decay to $l^+l^{\prime-}\nu_l\bar{\nu}_{l'}$ through
a pair of W bosons, see \eqref{proc2}. 
As we can see from the
figure, such $s$-channel resonance processes are even more sensitive to the influence of
ISR.  Indeed, the Higgs mass is not properly
reconstructed by the MEM likelihood fit even if ISR jets with $p_{\rm
T} > 40$~GeV are vetoed---it is only when the veto threshold is lowered to
an unrealistically low value of 6~GeV that the fit becomes marginally
consistent with the correct input value $m_h=180$~GeV.  On the other
hand, the purely kinematical boost correction (without reweighting)
already yields a fit result that is very close to $m_h=180$~GeV, while inclusion of
the Sudakov reweighting leads to near-perfect agreement with the input value.

This strong sensitivity of the MEM fit to ISR is a particular feature of
processes with narrow $s$-channel resonances, where  $\Gamma_{\rm res} \ll
m_{\rm res}$ [$e.\,g.$ process \eqref{proc2} with $m_h=180$~GeV and
$\Gamma_h\approx 0.6$~GeV].
When ignoring ISR with transverse momenta of a few GeV, it can become impossible
to find final-state momentum configurations where both the W boson and the Higgs
propagators are on-shell for a given event, since the magnitude of the momentum
mismeasurement is larger than the Higgs width.
Many events will therefore be poorly or wrongly reconstructed.
In contrast, a pair production process without an  $s$-channel resonance 
is much less sensitive to the precise value of the incident momenta $p_{\rm
in}^{(\prime)}$.

\vspace{\medskipamount}
While the application of the boost correction method is straightforward at the parton
level, 
the situation becomes more complex in a more realistic framework with
jet fragmentation and hadronization, as will be shown in the next section.


\section{Initial-State Radiation at Hadron Level}
\label{sc:hadron}

In this section the influence of ISR is investigated in a setup
including parton showering, hadronization and a simple detector simulation in
the event generation. As before, events have been generated with \textsc{Pythia
6.4} \cite{pythia}, but now using fully hadronized
events and including underlying event from additional parton interactions.
These events have been passed through the fast detector simulation \textsc{Pgs
4} \cite{pgs} with general LHC detector parameters. 
For top quark pair production with di-leptonic decay
(eq.~(\ref{proc})), 
all events in the sample are
required to contain two reconstructed leptons and 
at least two reconstructed jets with $p_{\rm T} > 50$~GeV,
but no other selection cuts have been applied besides the intrinsic
detector acceptance.
The acceptance function in  eq.~\eqref{logl} has been determined by
producing one million events with \textsc{Pythia/Pgs} for each $m_t$ value in
the fit, and counting the number of events that pass the selection cut. To
account for jet smearing, a double-Gaussian transfer function (see
eq.~\eqref{mem2}) has been included for the jet energies, where the parameters
of the double-Gaussian function have been tuned to a large sample of $t\bar{t}$
Monte-Carlo events generated with \textsc{Pythia} and passed through
\textsc{Pgs}.\footnote{ISR has been turned off for the generation of these
events, since it otherwise would affect the fit.}
For the lepton energies and all angular variables, simple delta
functions have been used in lieu of transfer functions.

For events with
more than two jets, we assume that the extra jets come from ISR.
In principle, heavy flavor tagging could be used to discriminate between the $b$
jets from top decay and the ISR, which is comprised mostly of light-quark jets.
Since the  $b$ jet identification is not unique, one would need to take into
account the appropriate $b$-tagging efficiencies. However, this possibility is
highly process-specific and would not work for processes where the heavy
particles decay into light jets only, and we therefore choose not to
exploit it here. 
For the sake of generality, we take the most conservative approach,
and consider all permutations of the jets in the event, irrespective
of their flavor content, as candidates to come from
the top quark decay\footnote{If an event has more than four jets we only
permute the four hardest jets to save computing time.}. 
The remaining jets are interpreted as stemming from
ISR, where all jets reconstructed by \textsc{Pgs} are included, without 
applying any additional cuts. 
The likelihoods for the permutations are added to form the event
likelihood. Although most of these permutations correspond to incorrect
jet-parton assignments, the amount of noise introduced into the fit is
negligible  since the wrong jet assignments typically result in very small
likelihood values.

\begin{figure}
\centering
\epsfig{figure=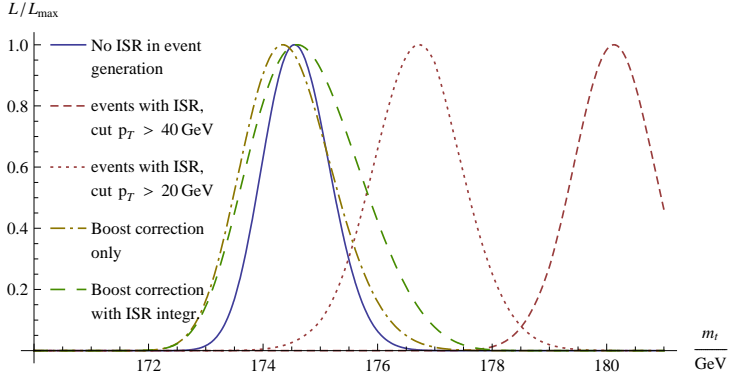, width=9.6cm}
\caption{Reconstruction of the top quark mass from a matrix
  element likelihood fit to 1000 hadron-level di-lepton $t\bar{t}$ events at the
  LHC with $\sqrt{s} = 14$~TeV. A top mass of $m_t = 175$~GeV has been used for
  the event generation. The solid curve corresponds to the idealized situation 
  without ISR in the event generation, while the result for uncorrected ISR is
  shown for a veto on extra jets with $p_{\rm T} > 40$~GeV (short dashed) and
  $p_{\rm T} > 20$~GeV (dotted). Also shown are the effect of the purely
  kinematical boost correction (dash-dotted) and the boost correction with ISR
  transfer functions (long dashed).
  The likelihood reflects statistical errors only.}
\label{hadron}
\end{figure}

Fig.~\ref{hadron} shows how ISR can affect the likelihood fit in this
more realistic case, if it is not accounted for in the matrix element
treatment. The solid curve again shows the situation without ISR in
the event generation\footnote{Even in the case without any ISR, 
the likelihood can be subject to a small bias caused by final-state
radiation. This bias can in principle be eliminated by correcting the
transfer functions for the effect of final-state
radiation; see also footnote \ref{fsrtrans} on page \ref{fsrtrans2}.}. 
The short-dashed curve corresponds to inclusion
of ISR in the event generation but not in the MEM likelihood fit, and
events with more than two jets with $p_{\rm T} > 40$~GeV have been
excluded from the fit. In spite of this cut, the central value for the
fitted top mass is shifted significantly compared to the input value
$m_t = 175$~GeV, a feature that was not observed in the parton-level
analysis (see Fig.~\ref{parton}). It can be explained by the fact that
ISR typically generates multiple jets per events, and even if each jet
has $p_{\rm T} < 40$~GeV, the total transverse ISR momentum can be
considerably larger. In events with very hard ISR, the kinematics of the 
final-state particles are substantially modified, so that the phase-space
integration for matrix elements without ISR correction is sometimes
pushed into an unphysical region. In such a situation, the fit results
can depend on the details of the computer implementation of the MEM, and do
not yield any meaningful information.

The situation improves somewhat when the jet $p_{\rm T}$ cut is
lowered to 20~GeV (dotted curve in Fig.~\ref{hadron}), but the
fit result is still clearly inconsistent with the input value $m_t =
175$~GeV. Moreover, such a low $p_{\rm T}$ cut will be subject to
large systematic experimental uncertainties.

It is therefore necessary to take into account and correct for ISR in
the MEM fit. In order to perform the boost correction described in
Section~\ref{sc:parton}, each ISR jet needs to be associated with one
of the incoming legs. A simple rule is to assume that jets in the left
hemisphere stem from the incident parton coming from the right, and
\emph{vice versa}. As a first step, we will not include
resolution functions for the ISR (in contrast to the other jets), in order to
minimize the computing time, but we will comment on their r\^ole later.
Similar to the parton-level analysis, the
application of the kinematical boost correction (dash-dotted curve)
leads to a considerably better agreement with the input value for
the top mass.

\begin{figure}
\centering
\epsfig{figure=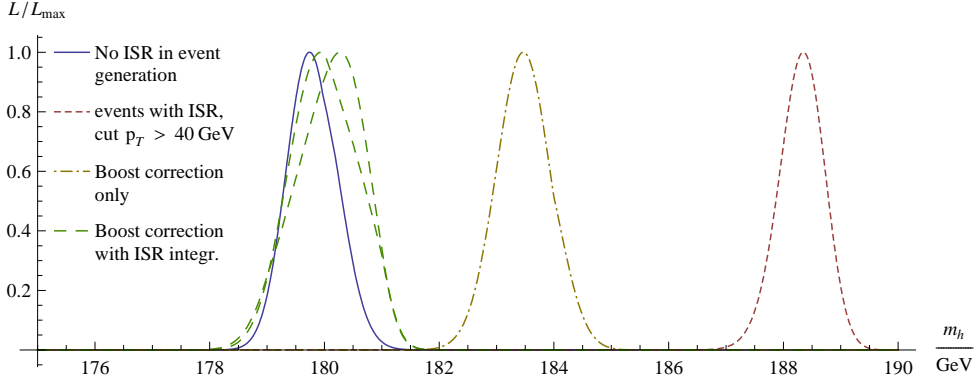, width=12.9cm}
\caption{Reconstruction of the Higgs boson mass from a matrix
  element likelihood fit to 1000 hadron-level di-lepton events at the LHC with
  $\sqrt{s} = 14$~TeV. A Higgs mass of $m_h = 180$~GeV has been used for the event
  generation. 
  The different curve correspond to the following setups:
  idealized situation without ISR in
  the event generation (solid curve); 
  ISR included in the simulation but no correction (short dashed);
  purely kinematical boost correction (dash-dotted);
  boost correction with ISR transfer functions (long dashed).
  The two long-dashed curves correspond to ISR transfer functions tuned to 
  $t\bar{t}$ and $h\to WW$ Monte-Carlo events, respectively.
  The likelihood reflects statistical errors only.}
\label{hadron2}
\end{figure}

As evident from the figure, the purely kinematical boost correction
already leads to a satisfactory likelihood fit for top quark pair
production events. The situation is different, however, for processes
with a narrow $s$-channel resonance, like Higgs production
\eqref{proc2}. We have seen already at parton level that this class of
processes is very sensitive to ISR. Numerical results for the MEM fit
are shown in Fig.~\ref{hadron2}, which shows that the boost correction
does not lead to a good fit. This can be explained by the fact that on
average the measured ISR jet momenta do not agree sufficiently well
with the parton-level ISR momenta.  Such measurement inaccuracies have
a substantial impact for Higgs production process due to its strong
sensitivity on the $p_{\rm T}$ of ISR.

While inclusion of Sudakov corrections was successful in the
pure parton level case of Section~\ref{sc:parton}, it turns out to be
less useful in the fully hadronic case, and hardly
improves the results from the pure boost correction. The reason for
this is that the imperfect reconstruction of ISR in the detector
has a much bigger impact on the likelihood fit. 

There is however another way to account for these strong ISR effects,
which drastically reduces the dependence on the detector
acceptance. By including a transfer function for the transverse
momentum of each incident particle (in addition to the transfer
functions for the outgoing legs of the matrix element), we can
successfully account not only for ISR that is visible in the detector,
but also for the case when the ISR does not produce visible jets.
We use a two-component transfer function, employing a double-Gaussian if the
measured ISR $p_{\rm T}^{\rm vis}$ is non-zero, and a single Gaussian in log-space for
zero $p_{\rm T}^{\rm vis}$:
\begin{align}
W_{\rm ISR}(p_{\rm T},p_{\rm T}^{\rm vis})
&= \left\{
\begin{array}{ll}
\frac{1}{\sqrt{2\pi}(a_2+a_3a_5)}
\bigl[ e^{-(p_{\rm T}-p_{\rm T}^{\rm vis}-a_1)^2/(2a_2^2)} + a_3\,
  e^{-(p_{\rm T}-p_{\rm T}^{\rm vis}-a_4)^2/(2a_5^2)}\bigr], & 
  \text{for } p_{\rm T}^{\rm vis} > p_{\rm T}^0, \\[1ex]
\frac{1}{\sqrt{\pi}\,b_2\,p_{\rm T}} \, e^{-(\log(p_{\rm T})-b_1)^2/(2b_2^2)}
& 
  \text{for } p_{\rm T}^{\rm vis} < p_{\rm T}^0,
\end{array}
\right. \nonumber \\[1ex]
& \quad \text{with } a_i = b_{i0} + b_{i1}\sqrt{p_{\rm T}} + b_{i2}p_{\rm T}.
\label{isrtrans}
\end{align}
The boundary $p_{\rm T}^0$ between the two regions should be chosen
near the sensitivity limit of the detector (typically a few GeV), 
but we have checked that the results are not
appreciably affected by varying $p_{\rm T}^0$ between 5 and 15~GeV.

The free parameters $b_i,\,b_{ij}$ 
in \eqref{isrtrans} are tuned to Monte Carlo simulated data, and it
has been checked that the transfer function provides a good approximation to the
Monte Carlo data both for small (a few GeV) and large ($\sim 100$~GeV) values of
$p_{\rm T}$. This tune effectively accounts for Sudakov factors, as
well as detector acceptance effects.
When using ISR transfer functions one needs to integrate over
the partonic $p_{\rm T}$ of each leg, so that the total integration dimension is
increased by two. Nevertheless, when using an adaptive algorithm
like \textsc{Vegas} \cite{vegas}, the integration time grows only by a factor of
less than 10. 

The increase in the number of degrees of freedom also leads to an
increase of the width of the curve---however, the expected reduction
in systematic effects and stability of the likelihood result due to
the better control of QCD radiation using this method should by far
make up for this.

Fig.~\ref{hadron2} demonstrates the
effect of the boost correction without and with ISR transfer functions
for the Higgs production process \eqref{proc2}.
The plot shows that the ISR transfer functions properly take into account the 
typical energy resolution
and jet smearing effects and the fit result is consistent
with the input value $m_h = 180$~GeV.

Note that Fig.~\ref{hadron2} shows two curves for the result with ISR transfer functions, which
correspond to transfer functions tuned to $t\bar{t}$
and $h\to WW$ Monte-Carlo events, respectively. The likelihood curves are
almost identical for the two cases,
which demonstrates that the transfer functions are quite insensitive to the hard
process.

For the case of top quark pair production, the fit result for the boost
correction with ISR transfer functions is shown by the long-dashed curve in
Fig.~\ref{hadron}. It agrees very well with the input value $m_t = 175$~GeV, but
in contrast to Higgs production, the purely kinematical boost correction is
already satisfactory so that the inclusion of the transfer functions does not
significantly improve the results.

\vspace{\medskipamount}
Since the integration over ISR transfer functions helps to improve the
MEM likelihood fit, one could wonder whether it might be sufficient to ignore
the measured ISR momenta altogether and instead simply integrate over the
$p_{\rm T}$ of ISR, weighted by the second line of eq.~\eqref{isrtrans}. We have
tested this idea, but found that the results of such a fit are
inconsistent with the true input values of $m_t$ or $m_h$, respectively. Similar
to the corresponding case without integration over $p_{\rm T}$ (dashed curves in
Figs.~\ref{hadron} and \ref{hadron2}), the best-fit values come out too
large. This
indicates that it is still important to include information about the observed QCD
radiation jets into the fit, even if this information is distorted by detector
effects.

\vspace{\medskipamount}
So far
we have shown that the boost correction method with ISR transfer functions is a
robust and practical technique for dealing with initial-state QCD radiation in 
experimental likelihood fits based on the MEM. It remains to check how our
method is affected by systematic uncertainties. The largest systematic
uncertainty is expected to be related to the jet energy scale. This
uncertainty can be taken into account by keeping the jet energy scale as a
free parameter in the fit \cite{toptev}.
However, an extensive determination of the statistical and systematic
uncertainties relating to the different methods (which are expected to
be larger than the purely statistical uncertainties given by the
likelihood fits in the figures), using
pseudo-experiments with varying input masses and simulation
parameters, is beyond the scope of this paper.

We will here instead focus on theoretical systematic error sources.
As already mentioned above, 
we have checked that the variation of the lower $p_{\rm T}$ cutoff for ISR
jets within reasonable ranges has a negligible effect on the fit results.
Similarly, is has been shown that the ISR transfer functions are approximately
universal and depend very little on the details of the hard scattering process
(see Fig~\ref{hadron2}).

Furthermore, we estimate the systematic error stemming from the parton
distribution functions (PDFs) by comparing fit results for CTEQ6L1 and CTEQ6M
PDFs \cite{cteq}. Here we only modify the PDFs in the MEM fit, while in
both cases using the same event file and transfer functions, which have been
determined with CTEQ6L1 PDFs. We find a negligible difference between the
results for CTEQ6L1 and CTEQ6M PDFs and thus conclude that the systematic error
from this source is very small.

Finally, missing virtual loop corrections in the hard matrix element
are expected to have a very small influence on the detailed kinematics of
the events. Since the matrix element method is not making use of the
total cross section for the determination of quantities such a masses
and spins, such corrections are expected to have a negligible impact. 
We have explicitly checked this  by running the same analyses on events
generated using \textsc{Powheg} \cite{powheg}, with no significant difference
in the results. The reader should note, however, that in order to ensure a
well-defined meaning of quantities such as masses at sub-GeV precision, this
procedure should, in the long run, be developed to work at full NLO level.


\section{Conclusions}
\label{sc:concl}

The Matrix Element Method (MEM) is a powerful tool for analyzing
processes with invisible particles in the final state at hadron
colliders. For each experimental event, the MEM computes a likelihood
that this event agrees with a given theoretical process supplied in
the form of the corresponding squared matrix element. However, the MEM
uses matrix elements with a fixed number of external partons, making
it difficult to include the high-multiplicity initial-state radiation
(ISR) expected to be abundant at the LHC. In this paper, it has been
shown explicitly that initial-state QCD radiation cannot be ignored in
MEM fits, without risk of unstable and biased results. 
The simplest way to circumvent this problem, by
applying a veto on events with sizeable ISR, is not acceptable since
this cut would need to be so severe that the statistics of the signal
event sample would be significantly depleted.

We have proposed a method to include the effect of ISR by correcting the momenta
of the incident partons in the matrix element on an event-by-event basis.
Concretely, the incoming parton momenta are boosted by the transverse momenta of
the ISR. The effectiveness of the method has been demonstrated by carrying out a
MEM fit for two characteristic physics processes with invisible
particles in the final state, $pp \to t\bar{t} \to b\bar{b}
l^+l^-\nu\bar{\nu}$ and $pp \to h \to W^+W^- \to l^+l^-\nu\bar{\nu}$. 

As a first
step, simulated parton-level events were used to show that
this boost correction significantly improves the result of the likelihood fit,
such that the fitted masses of the top quark or Higgs boson are fully consistent
with the respective input values.
As a second step, we applied the boost correction method to a more
realistic situation with fully hadronized events that were sent
through a fast detector simulation. In this case, detector effects
will typically lead to a mismatch between the reconstructed and the
true transverse momenta of the ISR.
This difference can be taken into account by including transfer
functions for the incident particles into the likelihood fit. The
transfer functions parametrize the distribution of reconstructed
transverse momenta for a given partonic transverse momentum, as
obtained from Monte Carlo data. As a byproduct, the transfer functions
also effectively capture the effect of showering, Sudakov factors, and
hadronization.
We found that the boost correction method with transfer functions yields stable
MEM fit results in excellent agreement with the underlying input values.

The proposed method increases the computing
time for the likelihood fit only by a moderate amount (less than a factor of ten
in all situations that we have studied). Furthermore it is very robust under the
influence of theoretical systematic uncertainties.

In conclusion, our results demonstrate that the boost correction
method with ISR transfer functions is a simple and effective technique
for treating ISR in MEM likelihood fits. It is however important to
validate our findings in a realistic experimental simulation with a
proper treatment of experimental systematic effects.


\section*{Acknowledgements}

The authors are grateful to D.~Whiteson, K.~Cranmer and T.~Plehn for interesting
discussions and insights. O.~M.\ would also like to thank F.~Maltoni for his
constant support. This project was supported in part by the National Science
Foundation under grant no.\ PHY-0854782, the HEPTOOLS EU network through Marie
Curie program RTN MRTN-CT-2006-035505, by Belgian Technical and Cultural Affairs
through the Interuniversity Attraction Pole P6/11,  and by Computational
Resources on PittGrid (\texttt{www.pittgrid.pitt.edu}). The work of J.~A.\ was
supported by NCTS under grant number NSC 98--2119--M--002--001. J.~A.\ and
A.~F.\ are thankful to the Aspen Center for Physics, where this project was
initiated.


\end{document}